\documentstyle[preprint,aps,prd]{revtex}

\newcommand\dd{{\rm d}}\newcommand\ee{{\rm e}}\newcommand\ii{{\rm i}}
\newcommand\half{{\textstyle\frac12}}\def\fourth{{\textstyle\frac14}}
\newcommand\sixth{{\textstyle\frac16}}\newcommand\eighth{{\textstyle\frac18}}

\begin{document}

\title{From classical chaos to decoherence in Robertson-Walker 
cosmology}    

\author{Fernando C.\ Lombardo \footnote{Electronic address: 
lombardo@df.uba.ar} and Mario Castagnino 
\footnote{Electronic address: castagni@iafe.uba.ar}}

\address{{\it
Departamento de F\'\i sica, Facultad de Ciencias Exactas y Naturales\\ 
Universidad de Buenos Aires - Ciudad Universitaria, 
Pabell\' on I\\ 
1428 Buenos Aires, Argentina}}

\author{Luca Bombelli \footnote{Electronic address: 
bombelli@olemiss.edu}}

\address{{\it
Department of Physics, 108 Lewis Hall, University of Mississippi\\ 
University, MS 38677, USA}}

\maketitle

\begin{abstract}
We analyse the relationship between classical chaos and particle
creation in Robertson-Walker cosmological models with gravity coupled to a
scalar field.  Within our class of models chaos and particle production are
seen to arise in the same cases.  Particle production is viewed as the seed
of decoherence, which both enables the quantum to classical transition, and
ensures that the correspondence between the quantum and classically chaotic
models will be valid.
\end{abstract}

\newpage

\noindent Historically, physics has provided a description of the arrow
of time by means of the second law of thermodynamics: The entropy of the
Universe, or any isolated system, grows with time.  However, physicists
generally accept that this second law is not a fundamental law of nature;
local, microphysical laws are invariant under time inversions, while
temporal asymmetry manifests itself in macroscopic physics and arises from
a restriction on the boundary conditions that selects the state of the
universe among those that satisfy the dynamical laws \cite{penrose}. In fact,
a number of apparently different, well known arrows of time can be identified
in different areas of physics, which are not completely independent.  In this
paper we would like to propose one such relationship, between the dynamical
instabilities that act as seeds of chaos at the classical level, and the
quantum to classical transition coming from particle creation in a
cosmological scenario \cite{calma}.

In a classical theory already, chaos opens up ways of explaining the origin
of the arrow of time \cite{calzet}, and therefore, in a cosmological setting,
of improving our knowledge of cosmological statistical mechanics and
thermodynamics.  Furthermore, just as non-integrability and dynamical
instability are, at the classical level, the causes of chaos, they are also,
at the quantum level, a cause of the instability of quantum states and
particles.  The question therefore arises, whether classical chaos and
semiclassical particle production are related.  At the semiclassical level,
one can prove that decoherence and correlations, which cause the transition
to the classical regime, arise in the presence of dynamical
instability \cite{calz-hu}; stable and unstable scalar field quanta are created
during this process.  The study of this connection between particle creation
and chaos is therefore of great interest.  If decoherence does induce a
transition from quantum to classical, then it should be possible to utilize
it in the context of quantum chaos to establish a more straightforward
correspondence between the behavior of classically chaotic systems and their
quantum counterparts \cite{zurek1}.

The definition of chaos for quantum systems and its experimental
consequences are still unclear.  In previous works this study has been
performed in differents ways.  Cooper {it et al}.  
\cite{cooper} considered
the coupling between a quantum oscillator and a classical one, which leads 
a chaotic dynamical system when a critical value of the coupling and energy 
is reached. This model could be related to the zero
momentum part of the problem of pair production of charged scalar particles
by a strong external electric field, and it would contain interesting 
information about the coupling between classical and quantum systems. Zurek 
and Paz \cite{zurek-jpp} have shown the implications of the decoherence 
process for quantum chaos in a
quantum open system composed of a chaotic system coupled to an environment.
Decoherence destroys quantum interference, so the correspondence between
quantum and classical models is mantained. The connection between classical 
and quantum systems have beeen also discussed 
by Cooper {it et al}. \cite{cooper1} analyzing the relationship between 
chaos and the variational approximation.

With this goal in mind, we will study a cosmological metric coupled to a
scalar field, and show that the conditions that produce chaotic behaviour at
the classical level give semiclassical particle production. Particle creation
is directly related to decoherence, and we will comment in our conclusions
about the fact that decoherence is the key element in both the particle
production $\leftrightarrow$ classical chaos correspondence, and the quantum
$\rightarrow$ classical transition in our cosmological model studied here.
 Although the relation between classical chaos and decoherence could be 
very interesting to be raised in general cases (outside of our cosmological 
model studied at present), we will only refer to the Robertson-Walker 
cosmological case in the paper; new and more general results will be 
communicate in turn \cite{nos}. 

Let us consider a Robertson-Walker (RW) universe in conformal time gauge,
\begin{equation}
   \dd s^2 = a^2(t) [ -\dd t^2 + \dd\chi^2 + \sin^2\!\chi
   (\dd\theta^2 + \sin^2\!\theta\,\dd\varphi^2) ] \;, \label{Metric}
\end{equation}
with dynamics described by the Einstein-Hilbert action
\begin{equation}
   S_{\rm grav}[g] = {1\over{16\pi G}}\int \dd^4x ~\sqrt{-g} ~R\;,
\end{equation}
conformally coupled to a real scalar field of mass $\mu$, with action
given by
\begin{equation}
   S_{\rm matter}[\Phi,g] = -\half \int \dd^4x ~\sqrt{-g}
   ~ \left[g^{\alpha\beta}\nabla_\alpha \Phi \nabla_\beta \Phi
   + (\mu^2+\sixth\,R)\, \Phi^2\right].
\end{equation}
If we reparametrize the scalar field by $\Phi\mapsto\phi = \sqrt{4\pi G/3}
\,a\Phi$, we obtain from the above action and the metric in (\ref{Metric})
the Hamiltonian
\begin{equation}
   H(a,\phi;\pi,p)
   = \half\, \big[-(\pi^2+a^2)+(p^2+\phi^2)+\mu^2a^2\phi^2\big]\;.
   \label{hami}
\end{equation}
The system described by this Hamiltonian consists of two harmonic
oscillators (of which one is ``inverted," as is to be expected from any
degree of freedom related to the spatial volume element), coupled through
a term proportional to $\mu^2$. For $\mu^2=0$, the system is trivially
integrable; for nonvanishing $\mu^2$, the coupling term introduces chaos
in the model, as shown in previous numerical \cite{caosnum} and
analytical \cite{bom-yo} work.

Let us summarize briefly our earlier results on the classical chaos in this
model \cite{bom-yo}. To begin with, we may replace $p$ and $\phi$ by new
dynamical variables $j$ and $\varphi$, respectively, defined by
\begin{equation}
   \phi = \sqrt{{2j\over\omega}} \sin\varphi\;, \qquad
   p = \sqrt{2 \omega j} \cos\varphi\;,
\end{equation}
where $\omega = \sqrt{1+\mu^2a^2}$ is the instantaneous frequency of the
field. The Hamiltonian in terms of the new variables can be written as a
sum $H = {\hat H}_0 + \delta {\hat H}$, of an unperturbed Hamiltonian which
is obviously integrable, and a perturbation.  With a second canonical
transformation we may resolve the unperturbed dynamics \cite{bom-yo}. We
indicate the transformation by $(a,\varphi;P,j) \mapsto (\theta,\delta;k,j)$,
where
\begin{equation}
   k = {j-h_0 \over \sqrt{1 - \mu^2j}}\;,
\end{equation}
and the angle variables canonically conjugate to $k$ and $j$, respectively,
are
\begin{equation}
   \theta = \arctan \left(\sqrt{1-\mu^2j} ~a/P\right), \qquad
   \delta = \varphi - {{\mu^2aP} \over{4\,(1-\mu^2j)}}\;,
\end{equation}
the Hamiltonian $H = H_0 + \delta H$ can be written (for small $\mu^2$) as
\begin{equation}
   H_0(k,j) = j - k \sqrt{1 - \mu^2 j}\;, \label{free}
\end{equation}
and
\begin{eqnarray}
   \delta H &=& \fourth\,\epsilon kj \Big[\cos(2\theta+2\delta)
   - \cos(2\theta-2\delta) \Big]
   + \fourth\,\epsilon^2 \Big[ -{\textstyle\frac32}\,k^2j - \fourth\,kj^2 +
\nonumber \\
   &&+\ (2 k^2j + \fourth kj^2) \cos2\theta - \half\, k^2j \cos4\theta
   - \half\, k^2j \cos(2\delta-\pi/2) +
\nonumber \\
   &&+\ \fourth\, kj^2 \cos4\delta\ +\ k^2j \cos(2\theta-2\delta)
   - k^2j \cos(2\theta+2\delta) +
\nonumber \\
   &&-\ \eighth\, kj^2 \cos(2\theta+4\delta)
   - \eighth\, kj^2 \cos(2\theta-4\delta) +
\nonumber \\
   &&+\ {\textstyle\frac{\sqrt5}4} k^2j \cos(4\theta+2\delta+\psi)
   - {\textstyle\frac{\sqrt5}4} k^2j \cos(4\theta-2\delta+\psi)\Big]
   + O(\epsilon^3)\;,\qquad \label{pert}
\end{eqnarray}
where $\psi = - \arcsin(1/\sqrt{5})$, and the perturbation parameter
$\epsilon$ is to be identified with $\mu^2$ \cite{bom-yo}. The dynamics of
the unperturbed Hamiltonian is trivial, and gives a conditionally periodic
motion $\theta = \theta_0+\omega_kt$; $\delta = \delta_0 + \omega_jt$, with
frequencies
\begin{equation}
   \omega_k = {\partial H_0\over{\partial k}} = -\sqrt{1-\mu^2 j}\;,\qquad
   \omega_j = {\partial H_0\over{\partial j}} =
   {\mu^2 k + 2 \sqrt{1 - \mu^2 j} \over 2\sqrt{1-\mu^2 j}}\;.\label{freqs}
\end{equation}

The effect of the perturbation $\delta H$ is felt in particular on the
rational tori of the unperturbed dynamics, the ones where the motion
becomes periodic, because the frequencies associated with the two degrees
of freedom are related by the resonance condition
\begin{equation}
   n_0 ~ \omega_k + m_0 ~ \omega_j = 0\;, \label{res}
\end{equation}
for some integer numbers $n_0$ and $m_0$.  A calculation shows that the
leading order perturbation in the Fourier expansion (\ref{pert}) for $\delta
H$ which is resonant with a torus in the $H = 0$ constraint surface is the
second order term in $\epsilon$ with $(n_0,m_0) = (4,2)$. Following the
standard procedure to find how $V_{42}$ affects the local dynamics near the
resonant torus, we go over to a set of adapted canonical variables $(\gamma,
\delta; K, J)$, chosen so that one of the momenta will still be a constant of
the motion under the resonant perturbation:
\begin{equation}
   K = {{k - k_0}\over{n_0}} = \fourth\,(k-k_0)\;,\nonumber
\end{equation}
\begin{equation}
   \gamma = n_0 ~\theta + m_0 ~\delta +\psi_0
   = 4~\theta + 2~\delta + \psi_0\;,\nonumber
\end{equation}
\begin{equation}
   J = -{m_0\over{n_0}} k+j = -\half\,k + j\;,
\end{equation}
where $\psi_0$ is some arbitrary, fixed angle.

If we suppose that $K\ll1$ is a small increment of the variable $k$ around
the resonant value $k_0$, then we can expand the Hamiltonian, written in
the new variables, in powers of $K$, and study the dynamics generated by
the leading terms. We begin with the resonant part of the perturbation,
\begin{eqnarray}
   \delta H^{(2)}
   &=& \epsilon^2 V_{42}^{(2)}(k,j) \cos(4\theta+2\delta+\psi)
   + \hbox{(terms with different $n$ and $m$)} \nonumber \\
   &=& {\textstyle\frac{\sqrt5}{16}}\,\epsilon^2 k_0^2j
   \cos(\gamma+\psi-\psi_0)
   + \hbox{(terms with different $n$ and $m$)}\;.\qquad
\end{eqnarray}
Here, $j$ is to be thought of as $j(K =0,J)$, with $J$ arbitrary.
The perturbation depends only on $\gamma$ and not on $\delta$, $J$ is still a
constant of the motion; for simplicity we fix its value at the resonant one,
$J_0$. Then for the unperturbed Hamiltonian we obtain
\begin{equation}
   H_0(K,J_0) = H_0(k_0,j_0) + \half\,\tilde\Omega K^2 + O(K^3)\;,
\end{equation}
where $\tilde\Omega =
8\mu^2\,(1-\mu^2j_0)^{-1/2} + \mu^4k_0\,(1-\mu^2j_0)^{-3/2}$.  So, to lowest
order in $K$ and $\epsilon$, the $K$ dynamics near the resonant torus is
generated by
\begin{equation}
   H_{\rm loc}(\gamma,K) = H_0(k_0,j_0) + \half\,\tilde\Omega K^2
   + {\textstyle\frac{\sqrt5}{16}}\, \epsilon^2 k_0^2 j_0
   \cos(\gamma+\psi-\psi_0)\;,
\end{equation}
which is the Hamiltonian of a well-known system, the non-linear pendulum.

This system has homoclinic orbits, and it can be shown \cite{bom-yo} by means
of the Melnikov method that a stochastic layer forms in the vicinity of the
destroyed separatrix, which acts as a seed for chaos.  Work is in progress on
the calculation of the time scale for chaos to set in; this is an important
aspect, and will be discussed elsewhere.

To show the quantum manifestation of the instabilities that lead to the
onset of chaos in the classical model as described above, we will now consider
the corresponding semiclassical model, in which $\Phi$ is a massive quantum
scalar field on a classical curved background.  Integrating out the quantum
field, we will evaluate the closed time path (CTP) effective action, which
provides us with the non-equilibrium effects of quantum fluctuations over
the classical metric \cite{calz-hu2}, and can be written as
\begin{equation}
   \ee^{\ii S_{\rm eff}[g^+,g^-]}
   = N \ee^{\ii(S_{\rm grav}[g^+]-S_{\rm grav}[g^-])}
   \int {\cal D}\Phi^+ {\cal D}\Phi^- \ee^{\ii(S_{\rm matter}[\Phi^+,g^+]
   - S_{\rm matter}[\Phi^-,g^-])}. \label{ctpeff}
\end{equation}
From this effective action it is possible to obtain the real and causal
equation of motion, taking the functional variation of the action with
respect to the $g_{\mu\nu}^+$ metric, and then setting $g_{\mu\nu}^+ =
g_{\mu\nu}^-$. This CTP effective action has been explicitly
evaluated \cite{yo-mazzi}, using the expression of the Euclidean effective
action and the running of the coupling constants.

The CTP effective action is directly associated with the Decoherence
Functional of Gell-Mann and Hartle \cite{gellmann}, and can alternatively be
written in terms of the Bogolubov coefficients connecting the in and out
basis in each temporal branch. This fact implies that there is decoherence
if and only if there is particle creation during the field
evolution \cite{calz-hu}. Explicitly, integrating the right-hand side of Eq.\
(\ref{ctpeff}), we can write it as the Decoherence Functional
\begin{equation}
   {\cal D}[g^+,g^-] = \ee^{\ii(S_{\rm grav}[g^+]-S_{\rm grav}[g^-] +
   \Gamma [g^+,g^-])}\;, \label{decofunct}
\end{equation}
where $\Gamma$ is the influence action for the scalar field, with the
conformal factor being treated as an external field.

As the CTP effective action is independent of the out quantum state, we
have the freedom of choosing an out particle model. It is convenient
to choose a common out particle model for both evolutions (the Cauchy data
are the same, although the actual basis functions will be different). The
positive-frequency time dependent amplitude functions $\Phi_{\pm}$ for the
conformal model in each branch are related to those $F$ of the out model
by $\Phi_{\pm} = \alpha_{\pm} F + \beta_{\pm} F^{\star}$, where
$\alpha_{\pm}$  and $\beta_{\pm}$ are the Bogolubov coefficients in each
temporal branch,  obeying the normalization condition $\vert
\alpha_{\pm}\vert^2 - \vert \beta_{\pm}\vert^2 = 1$. The CTP effective
action in Eq.\ (\ref{decofunct}) is found to be \cite{calz-hu}
\begin{equation}
   \Gamma = {\ii\over2} \ln [\alpha_-\alpha_+^{\star} - \beta_-
   \beta_+^{\star}]\;. \label{gama}
\end{equation}
So, there is decoherence (Im $\Gamma > 0$) if and only if there is particle
creation in different amounts in the two temporal evolutions.

The scalar fields on a Robertson-Walker metric can be separated into modes
\begin{equation}
   \Phi(t,\vec x) = \sum\nolimits_{\vec k} \Phi_{\vec k}(t)
   \,\ee^{\ii\,\vec k\cdot\vec x}\;,
\end{equation}
where $\Phi_{\vec k}(t)$ are the amplitude functions of the $\vec k$-th mode.
Defining $\phi_{\vec k}(t):= a(t) \Phi_{\vec k}(t)$ we can write down the
wave equation for the $\vec k$-th mode,
\begin{equation}
   \ddot{\phi}_{\vec k}(t) + \big\{k^2 + [\mu^2 + (\xi-\sixth)\,R]
   \,a^2\big\}\,\phi_{\vec k}(t) = 0\;.
\end{equation}
(Notice that in terms of $\phi_{\vec k}$ the rescaled homogeneous field $\phi$
used in the classical model is $\phi = \sqrt{4\pi G/3}\,\phi_{\vec 0}$.)
For massless conformally coupled ($\xi= \sixth$) fields, $\phi_{\vec k}(t)$
admits the solutions
\begin{equation}
   \phi_{\vec k}(t) = A\,\ee^{\ii\Omega t} + B\,\ee^{-\ii \Omega t}\;,
\end{equation}
which are travelling waves in flat space. Since $\Omega = \vert \vec k \vert
=$ const, the positive and negative frequency components remain separated
and there is no particle production.  More generally, the wave equation for
each mode has a time-dependent natural frequency given by
\begin{equation}
   \Omega^2 = k^2 + [\mu^2 + (\xi - \sixth)\,R]\,a^2\;.
\end{equation}
The dynamics of the background, through $a$ and $R$, excites the negative
frequency modes. This is the same problem as the Schr\"odinger equation with
time-dependent potential $V(t) = [\mu^2 + (\xi-\sixth)\,R]\,a^2$, which can
induce backscattering of waves.  In terms of this potential, we can write the
Bogolubov coefficients up to lowest order in $V(t)$ as \cite{birrel}
\begin{eqnarray}
   &&\alpha_k = 1 + {\ii\over{2 \Omega}} \int_{-\infty}^{+\infty}
   V(t) \,\dd t\nonumber \\
   &&\beta_k = -{\ii\over{2\Omega}} \int_{-\infty}^{+\infty}
   V(t) \ee^{-\ii 2 \Omega t} \,\dd t\;.
\end{eqnarray}
The number of created particles in the $\vec k$-th mode can alternatively
be given in terms of $\dot{\phi}_{\vec k}$ and $\phi_{\vec k}$ by
\begin{equation}
   N_{\vec k}= \vert \beta_{\vec k}\vert^2 =
   {1\over2\Omega}\left[ \vert \dot{\phi}_{\vec k}\vert^2
   + \Omega^2 \vert \phi_{\vec k} \vert^2\right] - \half\;.
\end{equation}

In this semiclassical model, we have evaluated the Bogolubov coefficients
in a general mode decomposition. To make a real contact with the classical
part described above, we must evaluate each expression in the homogeneous,
$k=0$ mode contribution; we have written the Bogolubov coefficients and the
number of particles created in terms of different modes only to maintain
generality in the semiclassical treatment.

In the massless case $V(t)$ is proportional to $(\xi - \sixth)$, and
vanishes for conformal coupling. Therefore this term is present when the
quantum fields are massive and/or when the coupling is not conformal. This
is to be expected, since the imaginary part of the CTPEA is directly
associated to gravitational particle creation. For massless, conformally
coupled quantum fields, particle creation takes place only when spacetime
is not conformally flat. Therefore the only contribution to the imaginary
part of the CTPEA would be proportional to the square of the Weyl tensor in
a general case. Since our present case is a conformally flat example, there
is no particle creation when the conformally coupled field is massless.
When the fields are massive and/or non-conformally coupled, particle
creation takes place even if the Weyl tensor vanishes.  This is why an
additional contribution proportional to $R^2$ appears in the imaginary
part of the effective action for the non conformal case \cite{yo-mazzi}.

We have shown that in a classical Robertson-Walker metric conformally
coupled to a scalar matter field, the chaos appears when the field is
massive.  This is the same condition for particle creation to take place,
at the semiclassical level. To make this point more explicit, we will
show the ``relation'' between particle creation and the onset of chaos.

The Hamiltonian description of the cosmological particle creation is
restricted to the dynamics of a finite system of parametric oscillators,
where the Hamiltonian is simply \cite{kang}
\begin{equation}
   {\tilde H}(t)
   = \half \sum\nolimits_{\vec k}\, (p^2 + \Omega^2 \phi_{\vec k}^2)
   = \sum\nolimits_{\vec k}\, (N_{\vec k} + \half)\,\Omega\;.
\end{equation}
This is the field part of the Hamiltonian of Eq.\ (\ref{hami}) (there, it
was  evaluated in the $k=0$ mode contribution). One can identify $\vert
\phi_{\vec k} \vert^2$ and $\vert \dot{\phi}_{\vec k}\vert^2$ with the
canonical coordinates $\phi^2$ and $p^2$ respectively, the eigenvalue of
${\tilde H}$ being the energy ${\tilde E} = (N_{\vec k}+\half)\,\Omega$.
When $\mu = 0$, the zero mode Hamiltonian ${\tilde H} = 0$, and the total
Hamiltonian of Eq.\ (\ref{hami}) is integrable (an inverted harmonic
oscillator Hamiltonian).  Therefore the resonance condition Eq.\ (\ref{res})
can be explicitly written in terms of the number of created particles as
\begin{equation}
   {m_0\over{n_0}} = {{2\,[1-\mu^2\Omega_0\,(N_0 + \half)/\omega]^{3/2}}
   \over{2-\mu^2 \Omega_0\,(N_0 + \half})/\omega}\;,
\end{equation}
where $\Omega_0^2 = \mu^2 a^2$ and $N_0$ are the time dependent frequency
and the particle number in the zero mode contribution, respectively.
Therefore in the massless case (no particle creation) this condition is
only satisfied by $n_0 = m_0$ and $\omega_k =\omega_j = 1$, {\it i.e.}, these
frequencies are completely degenerate. This agrees with what we saw in the
classical model, where $\mu^2=0$ implies $\omega_k =\omega_j = 1$ from Eq.\
(\ref{freqs}), so the perturbative treatment is not applicable, but there is
no perturbation and chaos disappears in the massless case.

Decoherence is the key to understanding the relationship between the arrows
of time in cosmology.  In the context of quantum open systems, where the
metric is viewed as the ``system'' and the quantum fields as the
``environment,'' decoherence is produced by the continuous interaction between
system and environment.  The non-symmetric transfer of information from system
to environment is the origin of an entropy increase (in the sense of von
Neumann), because there is loss of information in the system, and of the time
asymmetry in cosmology, because growth of entropy, particle creation and
isotropization show a tendency towards equilibrium.  However, decoherence is
also a necessary condition for the quantum to classical transition.  In the
density matrix formulation, decoherence appears as the destruction of
interference terms and, in our model, as the transition from a pure to a mixed
state in the time evolution of the density matrix associated with the RW
metric; the interaction with the quantum modes of the scalar fields is the
origin of such a non-unitary evolution \cite{casta}.

In this paper we saw that, in the cosmological model we considered, particle
creation and decoherence are the effect of resonances between the evolutions
of the scale factor $a$ and the free massive field $\phi$, which is the origin
of the chaotic behaviour.  In the semiclassical treatment we used a general
coupling $\xi$ between the metric and the scalar field, and saw that a
non-conformal coupling effectively led to a time-dependent mass $M^2 = \mu^2
+ (\xi-\sixth)\,R$ for $\phi$, and particle production, even when $\mu^2 = 0$;
a perturbative analysis of the classical model for $\xi\ne\sixth$ would reveal
the presence of chaos, again even in the massless case; a similar conclusion
would be reached by analyzing a model with a $\lambda\phi^4$ self-interaction
term in the action.

\acknowledgments

\noindent
This work was supported by Universidad de Buenos Aires, CONICET and
Fundaci\'on Antorchas. We are grateful to D.F. Mazzitelli, G.B. Mindlin
and J.P. Paz for many useful discussions.

\end{document}